# Spontaneous Radiative Cooling to Enhance the Operational Stability of Perovskite Solar Cells via a Black-body-like Full Carbon Electrode

Bingcheng Yu [1,3, *], Jiangjian Shi [1, *], Yiming Li [1,3], Shan Tan [1,5], Yuqi Cui [1,5], Fanqi Meng [6], Huijue Wu [1], Yanhong Luo [1,4], Dongmei Li [1,4, †], Qingbo Meng [1,2,4, †]

[1] Key Laboratory for Renewable Energy, Institute of Physics, Chinese Academy of Sciences, Beijing Key Laboratory for New Energy Materials and Devices, Beijing 100190, China.

[2] Center of Materials Science and Optoelectronics Engineering, University of Chinese Academy of Sciences, Beijing 100049, China.

[3] Huairou Division, Institute of Physics, Chinese Academy of Sciences, Beijing 101400, China.

[4] Songshan Lake Materials Laboratory, Dongguan 523808, Guangdong, China

[5] School of Physics Science, University of Chinese Academy of Sciences, Beijing 100049, China

[6] Laboratory of New Ceramics and Fine Processing, School of Materials Science and Engineering, Tsinghua University, Beijing 100084, China.

*These authors contributed equally to this work.

†Corresponding author. Email: dmli@iphy.ac.cn (D.L.); qbmeng@iphy.ac.cn (Q.M.)

**Abstract:** Operational stability of perovskite solar cells is remarkably influenced by the device temperature, therefore, decreasing the interior temperature of the device is one of the most effective approaches to prolong the service life. Herein, we introduce the spontaneous radiative cooling effect into the perovskite solar cell and amplified this effect via functional structure design of a full-carbon electrode (F-CE). Firstly, with interface engineering, >19% and >23% power conversion efficiencies of F-CE based inorganic $CsPbI_3$ and hybrid perovskite solar cells have been achieved, respectively, both of which are the highest reported efficiencies based on carbon electrode and are comparative to the results for metal electrodes. Highly efficient thermal radiation of this F-CE can reduce the temperature of the operating cell by about 10 °C. Compared with the conventional metal electrode-based control cells, the operational stability of the above two types of cells have been significantly improved due to this cooling effect. Especially, the $CsPbI_3$ PSCs exhibited no efficiency degradation after 2000 hours of continuous operational tracking.

Solar cells are sensitive to the temperature, and the temperature enhancement will result in performance degradation *(1-6)*. The elevated temperature of an operating cell mainly comes from the non-photoelectric conversion of the absorbed solar energy *(7)*. For a cell having a power conversion efficiency (PCE) of 25%, the maximum heating power from the one sun AM 1.5 G illumination could reach 750 W m$^{-2}$. Under some extreme conditions, this heating power can make the cell temperature exceed 100°C *(8-9)*. This high temperature is a severe threat to the cell operational stability and also put forward rigorous demand to the encapsulation reliability *(10)*. Among solar cells, perovskite solar cells (PSCs) are much more sensitive to the elevated temperature *(11-16)*. Firstly, to the perovskite absorber materials, elevated temperature could cause the organic component volatilization *(11,12)*, accelerate ion (defect) migration and phase segregation *(13,14)*, and induce metastable crystal structure transition *(15,16)*. Secondly, elevated temperature could change or destroy the microstructural morphology, chemical doping and charge transporting ability of the organic hole transporting material *(17,18)*. Thirdly, elevated temperature could induce and accelerate the atom or component diffusion between different functional layers *(19,20)*.

Numerous works have been reported attempting to overcome this operational stability issue based on improving the heat resistance of the cell functional layers. For instance, dimension regulation, additive engineering, surface passivation, and process optimization routes are employed to suppress the decomposition or phase transition of perovskite absorber layers at high temperatures *(21-24)*. A series of new hole transport layer materials with higher thermal stability have also been developed for the cell *(25,26)*. Although the obvious progress has been achieved in the past few years, the stability performance of the PSC is still less satisfactory than commercialized inorganic solar cells.

Along with focusing on materials and interfacial stabilities, thermal management toward the complete cell started to be received attention. The primary concept of thermal management has been introduced to the PSC solar system by reducing the heating power and enhancing the heat dissipation, for example, doping higher thermal conductivity materials, optimizing device geometry and attaching heat spreaders *(27-29)*. Under thermal equilibrium conditions, compared to promoting interior thermal conduction, enhancing the heat dissipation of the cell terminals in fact plays a more critical role in cooling the whole cell. In some practical applications, passive liquid flow cooling or heat sink structures have been integrated into the photovoltaic systems for this purpose; however, these external cooling components would significantly increase the photovoltaic installation and maintenance cost *(30-32)*. Thus, it is still a challenge to actively cool the operating cell in a low-cost, easy scale-up and large-scale way.

Herein, we introduce the spontaneous radiative cooling effect into the PSC to enhance terminal heat dissipation of the cell. This effect is realized by functional structure design of a full-carbon electrode (F-CE) that simultaneously has high thermal emissivity and excellent electrical properties. The superior interfacial contact and charge transporting ability of the F-CEs contribute to >19% and >23% PCEs of inorganic $CsPbI_3$ and hybrid PSCs, respectively, both of which are the highest reported efficiencies based on carbon electrode and are comparative to the results for metal electrodes. The radiative cooling effect of the F-CEs reduced the temperature of the operating cell (AM 1.5 G, 1 sun) by about 10 °C. For operational stability test, no PCE degradation was observed in F-CE based $CsPbI_3$ cell under continuously tracking over 2000 hours. For low/high temperature-cycle test (-20/60 °C), the F-CE based $CsPbI_3$ cell can sustain 95% of the initial PCE after 100 cycles while the relative PCE of the Au electrode-based cell drops by >35%. These results suggest that the electrode thermal radiative cooling approach can provide a universal, convenient and low-cost solution to overcome the efficiency degradation of the cell induced by temperature elevation during the cell operation.

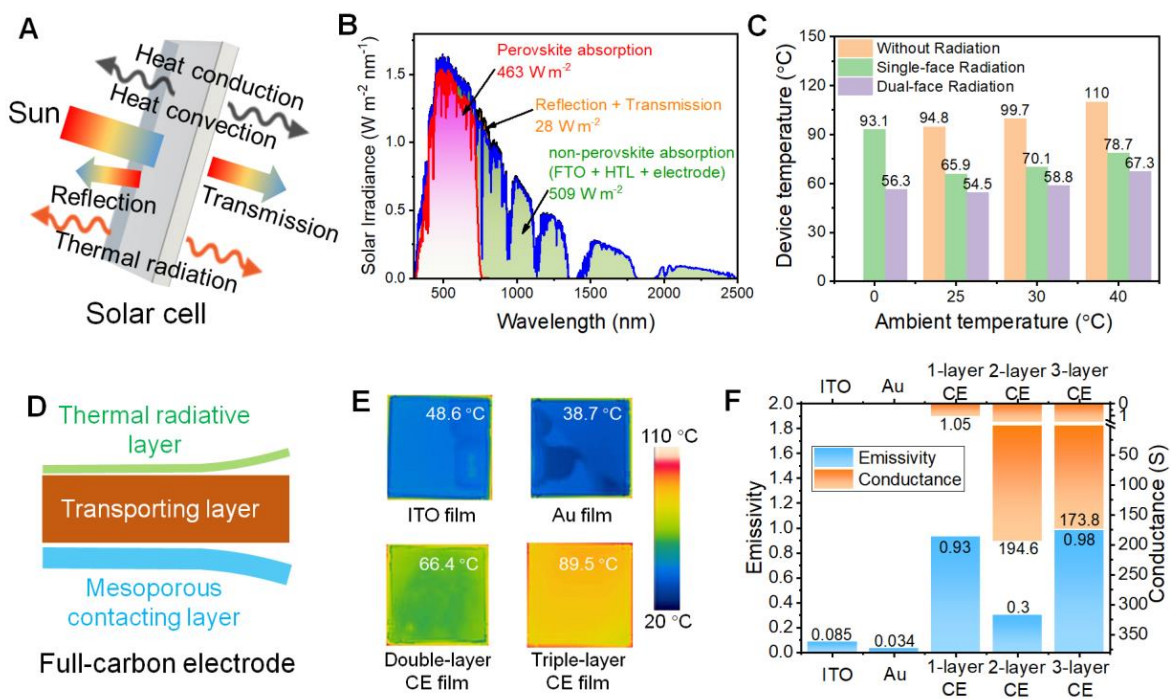

**FIG. 1 Thermal properties of the PSC and the radiative F-CE.** (A) Schematic diagram of the sun illumination and heat dissipation pathways of the cell. (B) Absorption of the AM 1.5 G solar irradiance spectrum by different parts of the cell, estimated from cell external quantum efficiency and reflection/transmission spectra. (C) Simulated cell temperature under heating power of 770 W m$^{-2}$. (D) Schematic diagram of the designed triple-layer F-CE comprising of mesoporous contacting, transporting and thermal radiative layers. (E) IR camera measured temperature of varied electrodes having surface temperature of about 90 °C. (F) Emissivity and conductance of the electrodes.

We first have a study on the heat generation and dissipation properties of the PSCs. As shown in FIG. 1A, thermal conduction, convection and radiation are the three main heat dissipation pathways. Heat generation properties caused by the light absorption of the PSC are experimentally estimated from light reflection/transmission and external quantum efficiency spectra of the cell (fig. S1-2). As in FIG. 1B, the cell absorbs about 97% of the entire sun illumination by the perovskite absorber layer (46%) and other functional layers (51%). Considering 20% PCE, the heating power of the cell illuminated under the AM 1.5G (1 sun) is 770 W m$^{-2}$. Using these parameters, the cell interior temperature was simulated by considering various thermal radiation configurations (fig. S3-6 and table S1). As presented in FIG. 1C, at room temperature, the interior temperature can reach 94.8°C if the cell does not have thermal radiative cooling. If the cell emits thermal radiation from a single surface, for example the conductive glass surface, the temperature obviously decreases to 65.9°C. This could be the practical condition of a PSC with metal back electrode. If the back electrode can also emit thermal radiation, the cell temperature will further decrease to 54.5°C, which can guarantee the cell working under a relatively moderate condition. This cooling effect of the dual-face thermal radiation can always work whatever the ambient temperature is. When the cell works in the vacuum, for example in the near space, the cell cooling effect benefited from the dual-face thermal radiation is more impressive (fig. S7).

Inspired by the simulation results, we have designed a triple-layer structural full-carbon film as the back electrode to realize dual-face thermal radiative cooling in the PSCs, as schematically shown in FIG. 1D and fig S8. This F-CE is comprised of three functional layers including a mesoporous layer having excellent contact with the hole transporting layer (HTL), a highly conductive graphite layer and a thermally radiative layer with efficient thermal radiation ability.

Thermal radiative properties of varied electrodes were characterized by comparing the IR apparent temperature $T_{IR}$ (measured with IR camera) and their real surface temperature $T_0$ (measured by thermocouple). As in FIG 1E, when increasing the $T_0$ to 90°C, $T_{IR}$ of the Au electrode is 38.7°C. For another alternative electrode, Sn: $In_2O_3$ (i.e., ITO), the $T_{IR}$ is 48.6°C. Comparatively, TIR of the triple-layer F-CE reaches 89.5°C, very close to the $T_0$. If the F-CE does not have the thermal radiative layer, its $T_{IR}$ is 66.4°C. Thermal radiation ability of these electrodes is further quantified by using emissivity ($\varepsilon$), that is, $\varepsilon \approx (T_{IR}/T_0)^4$ (33). The $\varepsilon$ of the Au and ITO electrode are only 0.034 and 0.085 (FIG. 1F), respectively, and for the simplest single-layer carbon electrode, its $\varepsilon$ is 0.93. If a graphite layer is introduced to enhance the carbon electrode conductance, the $\varepsilon$ decreases to 0.3 (fig. S8). For our designed triple-layer F-CE, the ε reaches 0.98, very close to an ideal black body. In addition, the triple-layer F-CE has large charge conductance because of the interior graphite layer, which is suitable for the PSCs (fig. S9).

We use the F-CE to fabricate inorganic $CsPbI_3$ and hybrid PSCs. The cell structure of the $CsPbI_3$ solar cell, comprised of FTO (F: $SnO_2$) glass, $TiO_2$, $CsPbI_3$, spiro-OMeTAD and F-CE, is shown in FIG. 2A. Carbon quantum dots (CQD) are introduced to improve the energy alignment of the F-CE/HTL interface. The CQDs were synthesized from hydrothermal method and evenly dispersed in water (FIG. 2B and fig. S10-11). These CQDs with the size of about 10 nm, can fill into the undulating area of the F-CE to reduce the surface roughness and modify the work function as well (fig S12-14). The F-CE was thermally pressed onto the top of the cell. This approach can ensure the F-CE robustly contact with the HTL with adhesive force reaching 3.6 N cm$^{-2}$, about two orders of magnitude higher than that of the HTL/Au contact (0.07 N cm$^{-2}$), as show in fig. S15. This mainly arises because the thermal-press approach can make the CE and HTL embed into each other, which will obviously change the film surface morphology and electrical potential (FIG. 2C and D). This not only benefits for the charge interface transfer but also facilitates the heat conduction within the cell.

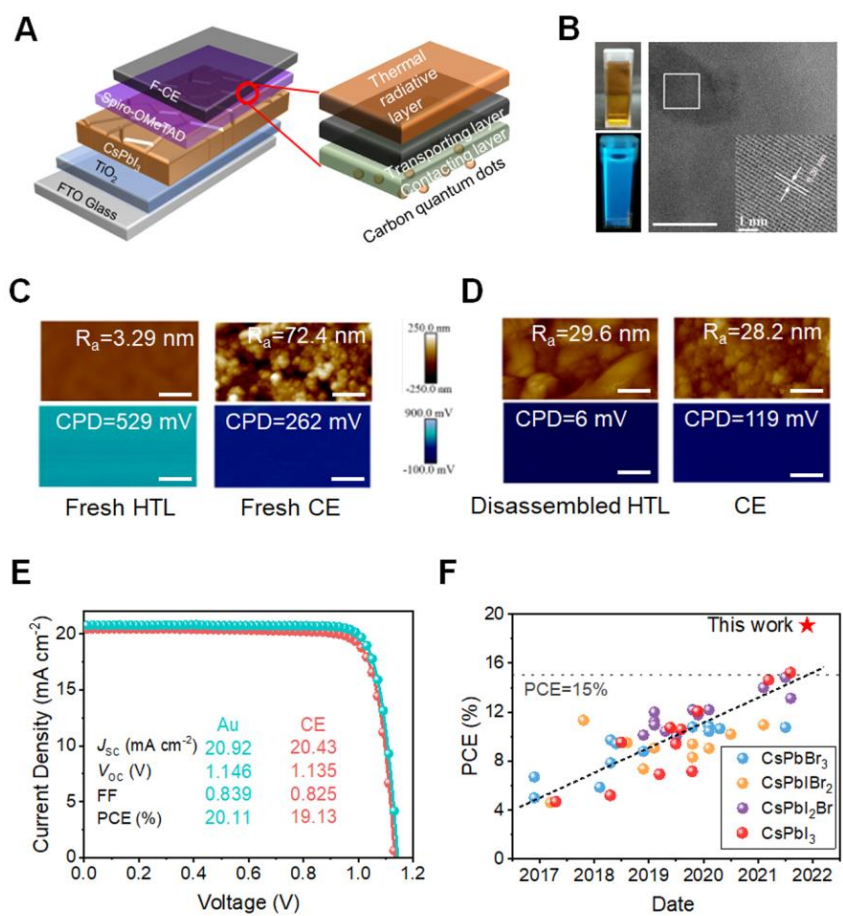

**FIG. 2 F-CE-based CsPbI$_3$ solar cell.** (A) Structure of the cell and the F-CE. Carbon quantum dots (CQD) are used to engineer the rear interface. (B) Photos of the CQD solution and its photoluminescence, and high-resolution transmission electron microscope image of the CQDs. Scale bar: 10 nm. Inset: lattice fringe image. (C-D) Surface morphology and contacting potential difference (CPD) of the fresh HTL and F-CE (C) and of them disassembled from the cell (D), respectively. Scale bar: 1 μm. (E) Current-voltage (*I-V*) characteristics of champion cells with Au electrode and F-CE, respectively. (F) PCE progress of the carbon electrode (CE) based inorganic CsPb(I, Br)$_3$ PSCs.

With the designed F-CE, a record 19.13% PCE (steady-state PCE: 18.7%) has been achieved for the CsPbI$_3$ cell with short-circuit current density ($J_{SC}$) of 20.43 mA cm$^{-2}$, open-circuit voltage ($V_{OC}$) of 1.135 V and fill factor (FF) of 0.825 (FIG. 2E, fig. S16-20, and table S2). This performance is comparative to the corresponding Au electrode-based cell (PCE: 20.11%). In the past few years, PCEs of the carbon-based inorganic perovskite cells (including CsPbI$_3$, CsPbI$_2$Br, CsPbIBr$_2$ and CsPbBr$_3$) indeed exhibited a linear increase tendency (FIG. 2F, table S3). As in mid-2021, the highest reported PCE was ~15% while the PCE gap between the Au and carbon-based cells exceeded 5%. Our result here has already narrowed the PCE gap to <1.0%. Besides the CsPbI$_3$ cells, we also achieved state-of-the-art carbon-based hybrid PSCs with 23.5% PCE (fig. S21-23). These results demonstrate that the F-CE is a promising electrode technology for the PSCs.

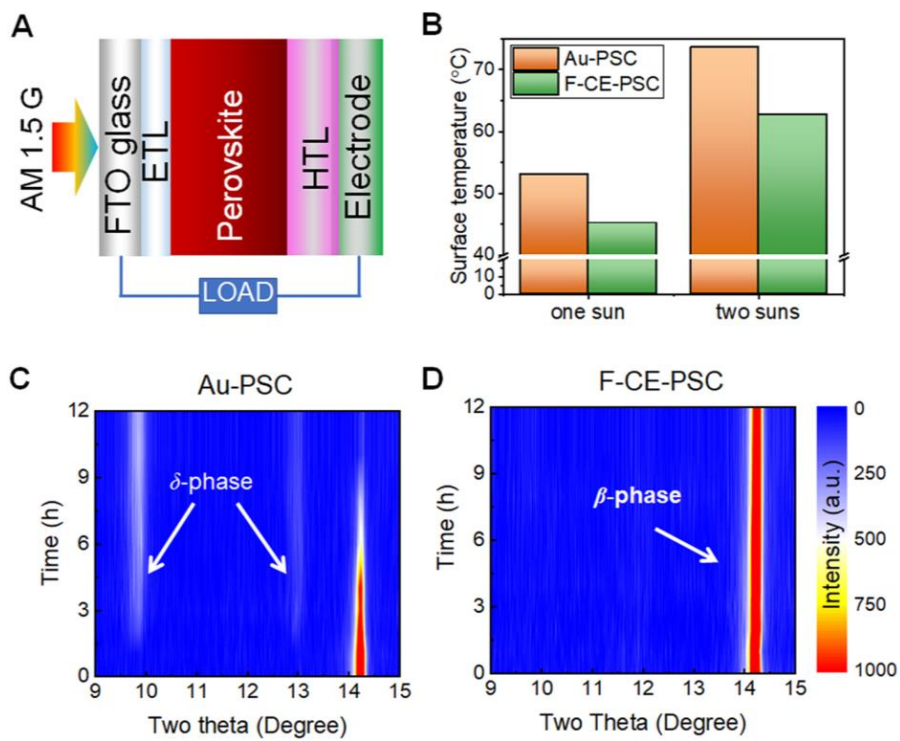

**FIG. 3 Influence of F-CE on the cell temperature and CsPbI$_3$ phase stability.** (A) Temperature measurement of an operating PSC under AM 1.5 G illumination. (B) Influence of electrode on the cell temperature. (C-D) Time-dependent X-ray diffraction patterns of the CsPbI$_3$ in the cells with varied electrodes under continuous light illumination (LED, about 6×10$^5$ lx).

We further experimentally evaluated the cooling effect of the F-CE on the temperature of the operating cell under AM 1.5 G illumination (FIG. 3A-B). Under one sun illumination, the F-CE can reduce the cell surface temperature by about 10°C, from 53.1 (Au electrode) to 45.2 °C; under two sun illumination, the cell temperature will reduce from 73.6 to 62.7 °C (fig. S24). This cooling effect will significantly enhance the ambient phase stability of the CsPbI$_3$ film in the cell. As indicated by time-dependent X-ray diffraction (XRD) shown in FIG. 3C-D, β-phase CsPbI$_3$ in the Au based PSC gradually transformed into δ-phase after being illuminated for several hours in ambient conditions. Comparatively, CsPbI$_3$ in the F-CE based PSC exhibited stable β-phase and constant XRD intensity in the whole illumination aging duration. It is also demonstrated that the temperature itself has obvious influence on the shelf life of the cell. Elevated temperature such as 65°C can easily cause PCE degradation and CsPbI$_3$ phase transition in a complete cell, whereas only 10°C temperature reduction will effectively improve the device stability (fig. S25-26).

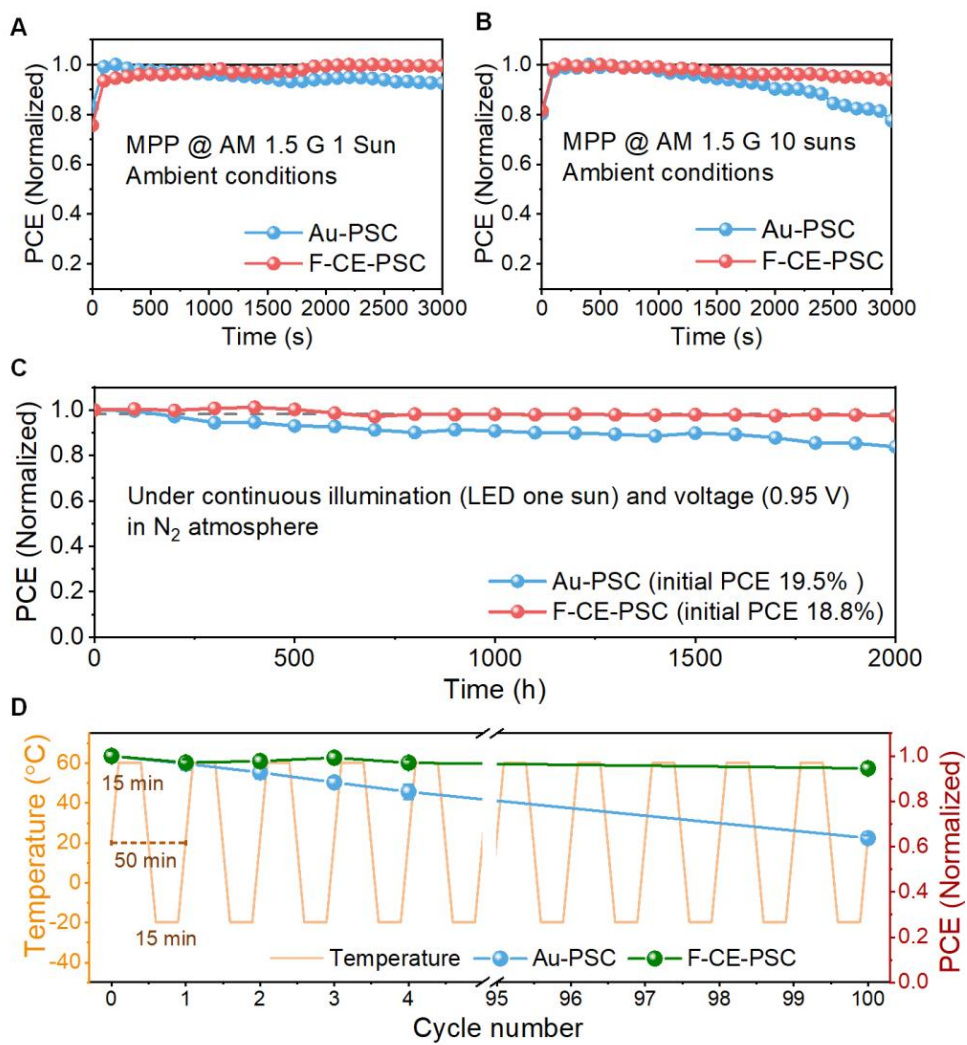

**FIG. 4 Operational stability test of the cell.** (A-B) PCE tracking of the cells operating under normal and 10× concentrated AM 1.5G illumination in ambient conditions. (C) 2000 h PCE tracking of the cells continuously operating under light illumination (initial cell current density 20

mA cm$^{-2}$) and bias voltage (0.95 V) in N$_2$ atmosphere. (D) Low (-20°C) -high (60°C) temperature cycling test in N$_2$ atmosphere for 100 cycles.

Lastly, we used different aging processes to evaluate the operational stability of the cells. After working at the maximum power point (MPP) under AM 1.5 G (one sun) for 3000 s in ambient conditions, no PCE degradation can be found for the F-CE based unencapsulated cell, whereas the PCE of the Au-based cell dropped to 92% of its initial value (FIG. 4A). Under concentrated 10 suns illumination, the PCE of the F-CE based cell only decreased to 94% of its initial value whereas the PCE of the Au electrode-based cell obviously dropped to 78% (FIG. 4B, and fig. S27-28). We further tracked 2000 hours' operational stability in N$_2$ atmosphere when keeping the cell continuously working under a steady-state bias voltage (0.95 V) and white LED illumination (initial cell current density 20 mA cm$^{-2}$). The PCE of the F-CE based cell slightly increased in the first 400 h, then kept almost constant from 500 to 2000 hours (FIG. 4C). To the best of our knowledge, this is one of the best operational stability results among the CsPbI$_3$ solar cells reported so far. Comparatively, the PCE of the Au based cell continuously degraded in the whole aging process, only sustaining 82% of its initial PCE after 2000 h. We also conducted low-high temperature (-20/60°C) aging test (FIG. 4D). The temperature range between -20 and 60°C was cycled for 100 times by using a semiconductor cooling stage, and the time duration at each temperature in one cycle is 15 min. After the aging process, the PCE of the Au-based cell has dropped to <70% of its initial value while the F-CE based cell still can sustain 95% of its initial value. This better temperature-cycling stability mainly benefits from both the temperature buffering effect of the F-CE and the robust HTL/F-CE interface contact (fig. S29).